\RequirePackage{snapshot}
\documentclass[10pt,preprint,numbers,nocopyrightspace,nodate]{sigplanconf}

\usepackage[table,usenames,dvipsnames]{xcolor}
\usepackage[T1]{fontenc}
\usepackage[tableposition=top]{caption}
\usepackage{rotating}
\usepackage{color, colortbl}
\usepackage{booktabs, array, dcolumn}
\usepackage{tikz}
\usepackage{graphicx}
\usepackage{pgf-umlsd}
\usepackage{multirow}
\usepackage{epsfig}
\usepackage{subfig}
\usepackage{pgfplotstable}
\usepackage{amsfonts}
\usepackage{relsize}
\usepackage{varwidth}
\usepackage{url}
\usepackage{amsmath}
\usepackage{amssymb}
\usepackage{booktabs}
\usepackage[normalem]{ulem}
\usepackage{subfig}
\usepackage{kernelstyle}
\definecolor{Gray}{gray}{0.9}
\usepackage{footnote}
\usepackage{longtable}
\usepackage{multirow}
\usepackage{siunitx}

\definecolor{gray1}{gray}{0.9}
\definecolor{gray2}{gray}{0.6}
\definecolor{Gray}{gray}{0.9}

\definecolor{light-gray1}{gray}{0.65}
\definecolor{light-gray2}{gray}{0.95}

\DeclareCaptionFormat{myformat}{%
  \begin
  {varwidth}{\linewidth}%
    \centering
    #1#2#3%
  \end{varwidth}%
}
\begin{document}
%
\title{A Neural Embeddings  Approach for Detecting Mobile Counterfeit Apps}

\authorinfo{Jathushan Rajasegaran}{Data61-CSIRO, Australia \\ University of Moratuwa, Sri Lanka}{}
\authorinfo{Suranga Seneviratne}{University of Sydney}{}
\authorinfo{Guillaume Jourjon}{Data61-CSIRO, Australia}{}

\maketitle

\begin{abstract}
Counterfeit apps impersonate existing popular apps in attempts to misguide users to install them for various reasons such as collecting personal information, spreading malware, or simply to increase their advertisement revenue. Many counterfeits can be identified once installed, however even a tech-savvy user may struggle to detect them before installation as app icons and descriptions can be quite similar to the original app. To this end, this paper proposes to use neural embeddings generated by state-of-the-art convolutional neural networks (CNNs)  to measure the similarity between images. Our results show that for the problem of counterfeit detection a novel approach of using style embeddings given by the Gram matrix of CNN filter responses outperforms baseline methods such as content embeddings and SIFT features. We show that further performance increases can be achieved by combining style embeddings with content embeddings. We present an analysis of approximately 1.2 million apps from Google Play Store and identify a set of potential counterfeits for top-1,000 apps. Under a conservative assumption, we were able to find 139 apps that contain malware in a set of 6,880 apps that showed high visual similarity to one of the top-1,000 apps in Google Play Store. 
\end{abstract}

\section{Introduction}
\label{Sec:Introduction}

Availability of third party apps is one of the major reasons behind the wide adoption of smartphones. The two most popular app markets, Google Play Store and Apple App Store, hosted approximately 3.5 million and 2.1 million apps at the first quarter of 2018~\cite{google2017,apple2017} and these numbers are likely to grow further as they have been in last few years. Handling such large numbers of apps is challenging for app market operators since there is always a trade-off between how much scrutiny is put into checking apps and encouraging developers by providing fast time-to-market. As a result, problematic apps of various kinds have made into the apps markets including malware, before they have been taken down after receiving user complaints~\cite{zhou2012dissecting,seneviratne2015early}. 

\begin{figure}[h]
\centering
\includegraphics[scale=0.3]{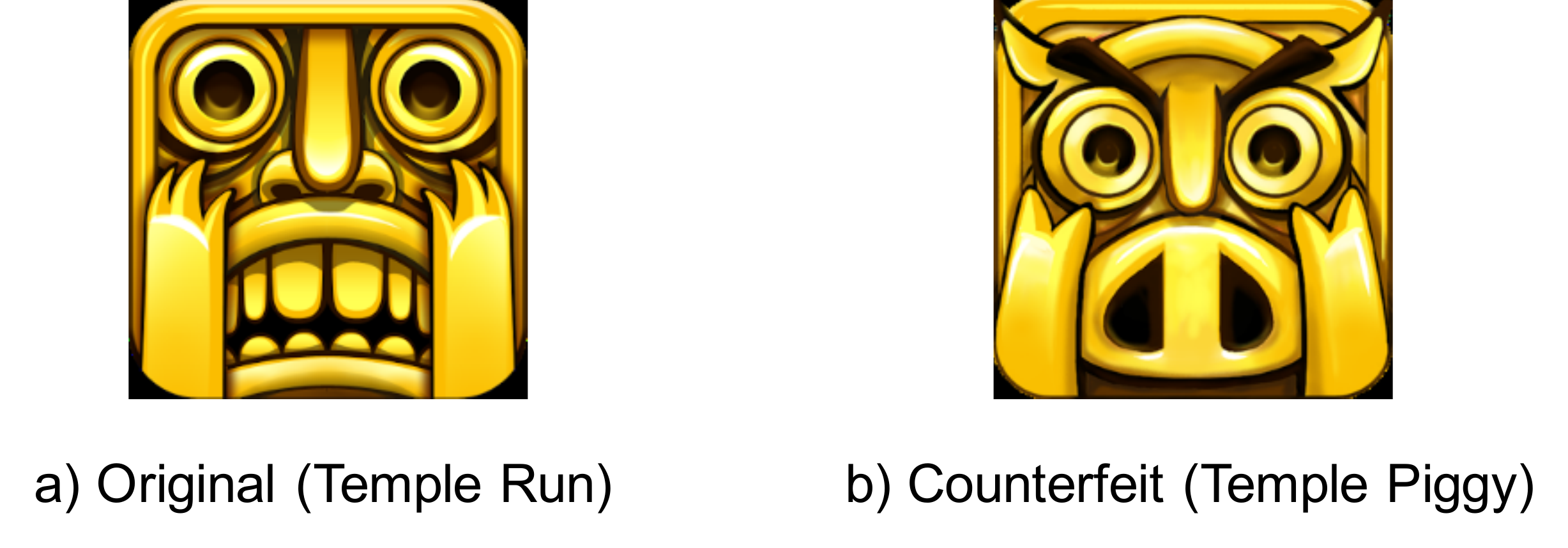}%
\caption{An example counterfeit app for the popular arcade game Temple Run} \vspace{-2mm}
\label{Fig:Ex1}
\end{figure}

One category of problematic apps making into app markets is \emph{counterfeits} (i.e. apps that attempt to impersonate popular apps). The overarching goals behind app impersonation can be broadly categorised into two. First, the developers of counterfeits are trying to attract app installations and increase their advertisement revenue. This is exacerbated by the fact that some popular apps are not available in some countries and users who search the names of those popular apps can become easy targets of impersonations. Second is to use counterfeits as a means of spreading malware. For instance, in November 2017 a fake version of the popular messenger app \emph{WhatsApp}~\cite{whatsapp2017} was able to get into Google Play Store and was downloaded over 1 million times before it was taken down. Similar instances were reported in the past for popular apps such as \emph{Netflix}, \emph{IFTTT}, and \emph{Angry Birds}~\cite{angrybirds2012,sarah2013new,netflix2017}. More recently, counterfeits have been used to secretly mine crypto currencies in smartphones~\cite{securelist2018}. In \figurename~\ref{Fig:Ex1} we show an example counterfeit named \emph{Temple Piggy}\footnote{\emph{Temple Piggy} is currently not available in Google Play Store.}  which shows a high visual similarity to the popular arcade game \emph{Temple Run}.\footnote{\emph{Temple Run} - https://play.google.com/store/apps/details?id=com.imangi.templerun.}


%
%
%


In this paper, we propose a neural embedding-based image retrieval framework that allows to identify visually similar apps to a given app icon from a large corpus of icons. Indeed, recent advances in Convolutional Neural Networks (CNNs) allow to generate feature embeddings to a given image that are related to the content of the image using pre-trained models such as AlexNet~\cite{alexNet}, VGGNet~\cite{vggNet}, and ResNet~\cite{resNet}. In particular, we demonstrate that content embeddings are moderately suitable for the task of visually similar app icon detection as many fake apps tend to have a similarity in the image style in terms of colour and texture than the actual content. In addition, we show that  \emph{style embeddings} generated from the Gram matrix of convolutional layer filter responses of the same pre-trained models achieve better results compared to baseline methods such as Scale Invariant Feature Transforms (SIFT)~\cite{SIFT} and combining content and style embeddings further increases the detection rates. More specifically, we make the following contributions:


\begin{itemize} 

\item We propose to use \emph{style embeddings} to detect visually similar mobile apps icons and introduce a novel image embedding that combines both content and style representations of an image that allows higher retrieval rates compared to baseline methods such as SIFT.

\item Using a large dataset of over 1.2 million app icons, we show that \emph{style embeddings} achieve 9\%-14\% higher retrieval rates in comparison to \emph{content embeddings} and SIFT-based methods.



\item We identify a set of 6,880 highly visually similar app icons to the top-1,000 apps in Google Play and show that under a conservative assumption, 139 of them contain malware.  To the best of our knowledge this is the first large-scale study that investigates the depth of app counterfeit problem in app stores.
\end{itemize} 

The remainder of the paper is organised as follows. In Section~\ref{Sec:RelatedWork}, we present the related work. In Section~\ref{Sec:Dataset}, we present our dataset followed by the methodology in Section~\ref{Sec:Methodology}. Section~\ref{Sec:Results} presents our results. We discuss limitations and possible future improvements in Section~\ref{Sec:Discussion} and Section~\ref{Sec:Conclusion} concludes the paper.

\section{Related Work}
\label{Sec:RelatedWork}

\subsection{Mobile Malware \& Greyware}
While there is a plethora of work on detecting mobile malware~\cite{grace2012riskranker,wu2012droidmat,burguera2011crowdroid,shabtai2012andromaly,yuan2014droid} and various fraudulent activities in app markets~\cite{xie2015appwatcher,chandy2012identifying,gibler2013adrob,surian2017app}, only a limited amount of work focused on the \emph{similarity of mobile apps}. One line of such work focused on detecting clones and rebranding. Viennot et al. ~\cite{viennot2014measurement} used the Jaccard similairty of app resources in the likes of images and layout XMLs to identify clusters of similar apps and then used the developer name and certificate information to differentiate clones from rebranding. Crussell et al.~\cite{crussell2013andarwin} proposed to use features generated from the source codes to identify similar apps and then used the developer information to isolate true clones. In contrast to above work, our work focuses on identifying visually similar apps rather than the exact similarity (i.e. clones), which is a more challenging problem.

Limited amount of work focused on identifying visually similar mobile apps~\cite{sun2015droideagle,malisa2016mobile,andow2016study,Malisa:2017}. For example, Sun et al.~\cite{sun2015droideagle} proposed DroidEagle that identifies the visually similar apps based on the LayoutTree of XML layouts of Android apps. While the results are interesting this method has several limitations. First, all visually similar apps may not be necessarily similar in XML layouts and it is necessary to consider the similarities in images. Second, app developers are starting to use code encryption methods, thus accessing codes and layout files may not always possible. Third, dependency of specific aspects related to one operating system will not allow to make comparisons between heterogeneous app markets and in such situations only metadata and image similarity are meaningful. Recently, Malisa et al.~\cite{Malisa:2017} studied how likely would users detect spoofing application using a complete rendering of the application itself. To do so, authors introduced a new metric representing the distance between the original app screenshot and the spoofing app. In contrast to above work, the proposed work intends to use different neural embeddings derived from app icons that will better capture visual similarities.



\subsection{Visual similarity \& style search}

Number of work looked into the possibility of transferring style of an image to another using neural networks. For example, Gatys et al.~\cite{gatys2015neural,gatys2016image} proposed a \emph{neural style transfer algorithm} that is able to transfer the stylistic features of well-known artworks to target images using feature representations learned by Convolutional Neural Networks (CNNs). Several other methods proposed to achieve the same objective either by updating pixels in the image iteratively or by optimising a generative model iteratively and producing the styled image through a single forward pass. A summary of the available style transfer algorithms can be found in the survey by Jing et al.~\cite{jing2017neural}.

Johnson et al.~\cite{johnson2016perceptual} have proposed a feed-forward network architecture capable of real-time style transfer by solving the optimisation problem formulated by Gatys et al.~\cite{gatys2016image}. Similarly, to style transfer, convolutional neural networks have been successfully used for image searching. In particular, Bell~\&~Bala~\cite{bell2015siamese} proposed a Siamese CNN to learn a high-quality embedding that represent visual similarity and demonstrated the utility of these embeddings on several visual search tasks such as searching products across or within categories. Tan et al.~\cite{tan2016ceci} and  Matsuo \& Yanai~\cite{style_classification} used embeddings created from CNNs to classify artistic styles. In contrast to these work, our work focuses on retrieving visually similar Android apps and we highlight the importance of style embeddings in this particular problem.


\section{Dataset}
\label{Sec:Dataset}

We collected our dataset by crawling Google Play Store using a Python crawler between January and March, 2018. The crawler was initially seeded with the web pages of the top free and paid apps as of January, 2018 and it recursively discovered apps by following the links in the seeded pages and the pages of subsequently discovered apps. Such links include apps by the same developer and similar apps as recommended by Google. For each app, we downloaded the metadata such as app name, app category, developer name, and number of downloads as well as the app icon in \emph{.jpg} or \emph{.png} format (of size 300 x 300 x 3 - height, width, and three layers for RGB colour channels). The app icon is the same icon visible in the smartphone once the app is installed and also what users see when browsing Google Play Store. 

We discovered and collected information from {\bf\emph{1,250,808 apps}} during this process.  For each app, we also downloaded the app executable in APK format using \emph{Google Play Downloader via Command line}\footnote{https://github.com/matlink/gplaycli} tool by simulating a Google Pixel virtual device. We were able to download APKs for {\bf\emph{1,024,511 apps}} out of the total {\bf\emph{1,250,808 apps}} apps we discovered. The main reason behind this difference is the \emph{paid apps} for which the APK can't be downloaded without paying. Also, there were some apps that did not  support the virtual device we used. Finally, the APK crawler was operating in a different thread than the main crawler as the APKs download is slower due to their large sizes. As a result, there were some apps that were discovered, yet by the time APK crawler reaches them they were no longer available in Google Play Store. \\  

\noindent{\bf \emph{Labelled set}}: To evaluate the performance of various image similarity metrics we require a ground truth dataset that contains similar images to a given image. We used a heuristic approach to shortlist a possible set of visually similar apps and then refined it by manual checking. Our heuristic is based on the fact that there are apps in the app market that have multiple legitimate versions. For example, popular game app \emph{Angry Birds} has multiple versions such as \emph{Angry Birds Rio}, \emph{Angry Birds Seasons}, and \emph{Angry Birds Go}. However, all these apps are using the same characters and as such icons are similar from both content (i.e. birds) and stylistic point of view (i.e. colour and texture). 

\begin{figure}
\centering
\includegraphics[trim=0cm 0cm 0cm 0cm, clip=true,scale=0.15]{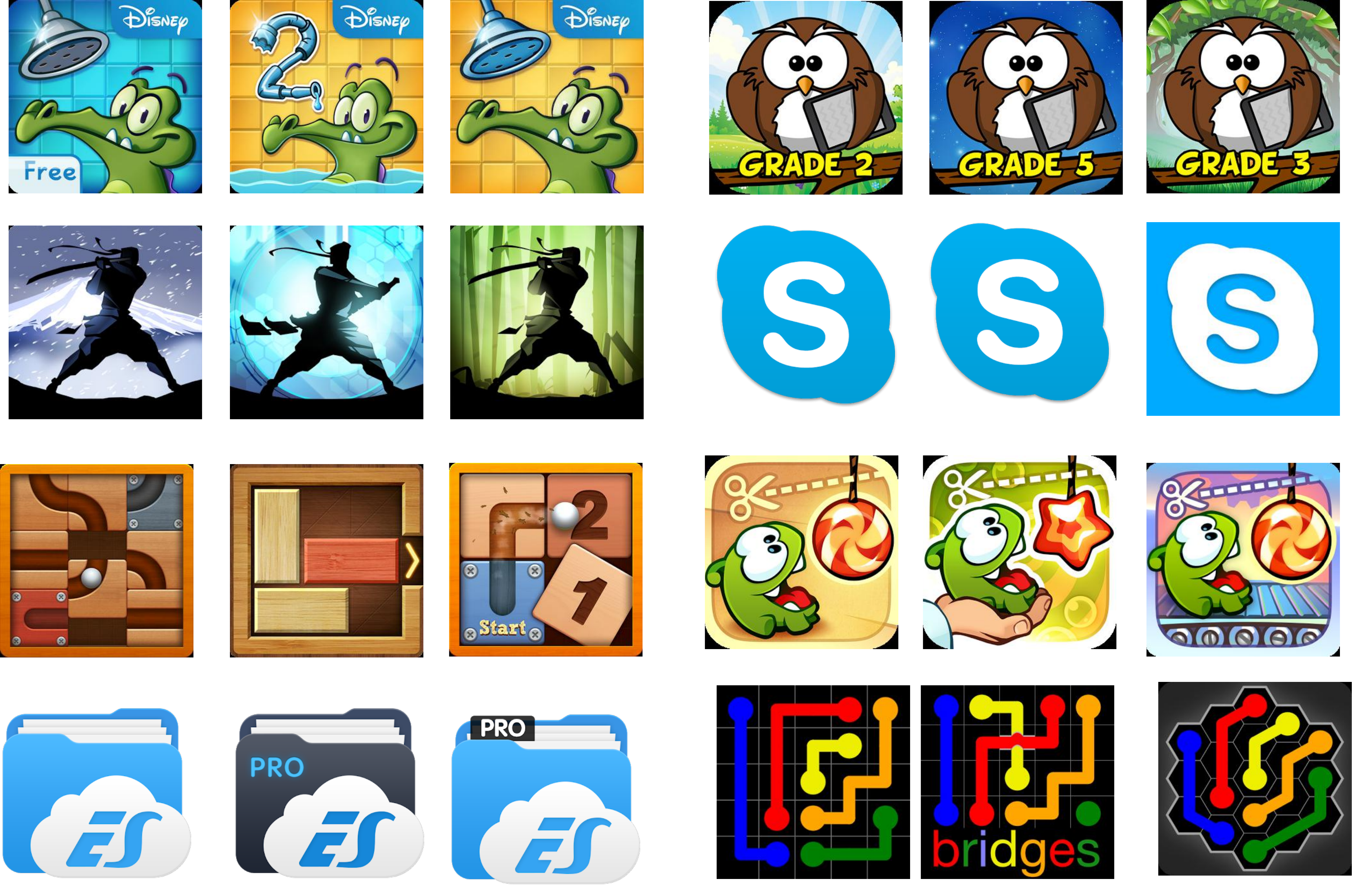}
\caption{Example similar app groups from the labelled set}\vspace{-2mm}
\label{Fig:label_dataset}
\end{figure}

Thus, we first identified the set of developers who has published more than two apps in app store and one app has at least 500,000 downloads. In a set of apps from the same developer, the app with the highest number of downloads was selected as the \emph{base app}. For each other app in the set, we then calculated the \emph{character level cosine similarity} of their \emph{app name} to the base app name and selected only the apps that had over 0.8 similarity and in the same Google Play Store \emph{app category} as the base app. Through this process we identified 2,689 groups of apps. Finally, we manually inspected each of these groups and checked whether the group consists of actual visually similar apps. In some occasions we found that some groups contained apps that are not visually similar and we discarded those groups. Also, we found that in some groups there were apps that were not visually similar to the others of the group and we removed those apps from the group. At the end of this process we had 806 app groups having a total of 3,910 apps as our {\bf \emph{labelled set}}. We show some example app groups from our {\bf \emph{labelled set}} in \figurename~\ref{Fig:label_dataset}.

\noindent{\bf \emph{Top-1,000 apps}}: To establish a set of potential counterfeits and to investigate the depth of app counterfeit problem in Google Play Store, we used top-1,000 apps since counterfeits majorly target popular apps. We selected top-1,000 apps by sorting the apps by the number of downloads, number of reviews, and average rating similar to what was done in our previous work~\cite{seneviratne2015early}. As we describe later in Section~\ref{Sec:Results}, for each app in top-1,000 we queried the $top-10$ similar apps by icon similarity in the set of all 1.2 million apps and checked whether the retrieved apps contain malware using the online malware check tool VirusTotal.\footnote{https://www.virustotal.com} 

\section{Methodology}
\label{Sec:Methodology}
As mentioned before, the main problem we are trying address is that \emph{``given an app icon how can we find visually similar icons from a large icon corpus?''}.
To solve this problem, our methodology involves two main phases; {\bf \emph{icon encoding}} and {\bf \emph{icon retrieval}} as discussed below.

\subsection{App Icon Encoding}
We encode the original app icon image of size $300 \times 300 \times 3$ to a lower dimension for efficient search as well as to avoid false positives happening due to Euclidean distance ($L_2$ distance) at large dimensions~\cite{L2_distance_not_good_for_high_diamentions}. We create three types of low dimensional representations of the images. \\ \vspace{-2mm}

\noindent{{\bf \emph{i) SIFT features}}:
As a baseline method, we use the \emph{Scale-Invariant Feature Transform (SIFT)}~\cite{SIFT} that extracts features from an image which are invariant to scale and rotation. SIFT features are commonly  used in object detection, face recognition, and image stitching. We generated lower dimensional SIFT representations for all 1.2 million apps. These representations are of dimension $t_i \times 128$,  where $t_i$ is the number of key-points detected by SIFT detector for icon $i$.  \\ 



\noindent{{\bf \emph{ii) Content embeddings}}: To extract the content representation of an icon we used the pre-trained \emph{VGGNet}~\cite{vggNet}; a state-of-the-art convolutional neural network that is trained on ImageNet~\cite{imagenet_cvpr09}. The fully connected layers of VGGNet constitutes a potential good representation of the content features of the app icon. We fed all 1.2M app icons to the VGGNet, and use the \emph{content embeddings}, $C \in \RR^{4096}$ generated at the last fully connected layer of VGGNet (usually called as the ${fc\_7}$ layer) that have shown good results in the past~\cite{fc7_is_good,cornell}.  \\ 


\noindent{{\bf \emph{iii) Style embeddings}}: As mentioned in Section~\ref{Sec:Introduction}, content similarity itself is not sufficient for counterfeit detection as sometimes developers keep the visual similarity and change the content. For example, if a developer is trying to create a fake app for a popular game that has birds as characters, they can create a game that has the same visual \emph{``look and feel''} and replace birds with cats. Therefore, we require an embedding that represent the style of an image.

Several work demonstrated that the filter responses of convolutional neural networks can be used to represent the style of an image~\cite{gatys2016image,gram}. For example, Gayts et al.~\cite{gatys2016image} used pre-trained convolutional neural networks to transfer the style characteristics of an arbitrary source image to an arbitrary target image. This was done by defining an objective function which captures the \emph{content loss} and the \emph{style loss}. To represent the style of an image, authors used the Gram matrix of the filter responses of the convolution layers. We followed a similar approach and used the fifth convolution layer (specifically $conv5\_1$) of the VGGNet to obtain the style representation of the image, as previous comparable work indicated that $conv5\_1$ provides better performance in classifying artistic styles~\cite{style_classification}. In the process of getting the embeddings for icons, each icon is passed through the VGGNet, and at $conv5\_1$ the icon is convolved with pre-trained filters and activated through $relu$ (Rectified Linear Unit) activation function.



More specifically, for an image $\image$, let $\Fl \in\RR^{\Nl\times\Ml}$ be
the filter response of layer $l$, where $\Nl$ denotes the number of filters in layer $l$
and $\Ml$ is the height times width of the feature map. $F^l_{ij}$ is the activation of $i^{th}$ filter at position $j$ in the layer $l$.

Similar to Gayts et al.~\cite{gatys2016image}, to capture style information we use the correlations of the activations calculated by the dot product. That is, for a given image $\image$, let $\Gl\in\RR^{\Nl\times\Nl}$ be the dot product Gram matrix at layer $l$, i.e.

\begin{equation}
   \Glij = \sum_{k=1}^{\Ml}\Flik\Fljk,
  \label{eqn:Glijdotproduct}
\end{equation}
where $\Fl\in\RR^{\Nl\times\Ml}$ is the activations of $I$. Then, $\Gl$ is used as the style representation of an image to retrieve similar images. The $conv5\_1-layer$ of the VGGNet we use has 512 filters  and thus the resulting Gram matrix is of size $G^5 \in \RR^{512 \times512}$. 




Gram matrix is symmetric as it represents the correlations between the filter outputs. Therefore, we only consider the upper triangular portion and the diagonal of the Gram matrix as our style representation vector, $S \in \RR^{131,328}$.
Though this reduces the dimension of the style vector by about half, the style embedding dimension is much larger compared to the content embeddings, $C \in \RR^{4,096}$. Thus, to further reduce the dimension of style embeddings we used \emph{very sparse random projection}~\cite{very_sparse_random_projection}. We selected sparse random projection over other dimensionality reduction methods such as PCA and t-SNE due to its computational efficiency.



More specifically, let $\mathbf{A} \in \RR^{n \times D}$ be our style matrix that contains the style embeddings of a mini batch of $n$ (in the experiments we used $n$=20,000) icons stacked vertically, and $D$ is the dimension of the style vector, which in this case is $131,328$. Then, we create a sparse random matrix $\mathbf{R} \in \RR^{D\times k}$ and multiply it with $A$. 
The elements $r_{ij}$ of $\mathbf{R}$ are drawn according to the below distribution,

\begin{equation} \label{eqn:sparse_matrix}
r_{ij} = \sqrt[4]{D}\left\{\begin{array}{lr}
1, & \text{with prob. } \frac{1}{2\sqrt{D}} \\
 0, & \text{with prob. } 1- \frac{1}{\sqrt{D}} \\
-1, & \text{with prob. } \frac{1}{2\sqrt{D}} \\
 \end{array}\right. 
\end{equation}


At large dimensions $\mathbf{R}$, becomes sparse as the probability of getting a zero is increasing with $D$. Since sparse matrices are almost orthogonal~\cite{sparse_random_projection_orthogonal}~\cite{sparse_random_projection_orthogonal2}, multiplication with $\mathbf{R}$, projects $\mathbf{A}$ in another orthogonal dimension.

\begin{equation}
   \mathbf{B} = \frac{1}{\sqrt{k}}\mathbf{A}\mathbf{R} \in \RR^{n \times k}
  \label{eqn:random_production}
\end{equation}

Each row of $\mathbf{B}$ gives a dimensionality reduced vector $S' \in \RR^{k}$ and in our case we used  $k=4,096$ to ensure the size of style embeddings matches the size of the content embeddings. In \figurename~\ref{Fig:Encoding}, we show a summary of our icon encoding process.

\begin{figure*}[ht!]
\centering
\includegraphics[trim=0cm 0cm 1cm 0cm, clip=true, scale=0.35]{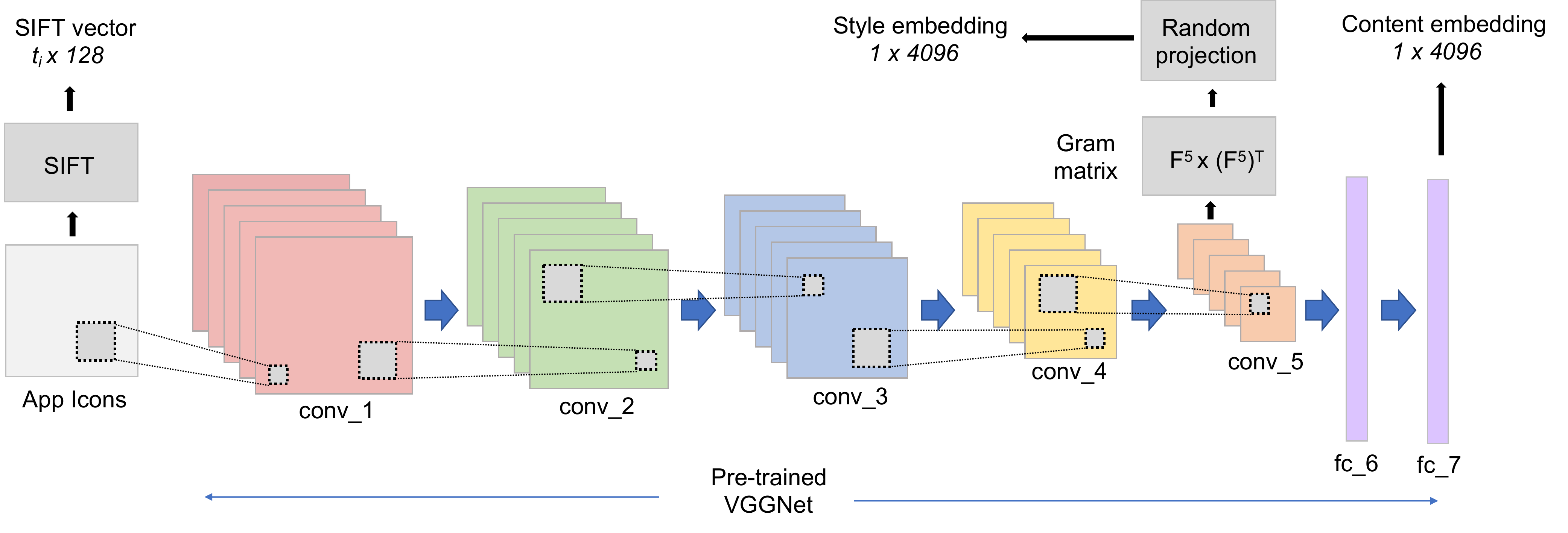}%
\caption{Summary of the icon encoding process}
\label{Fig:Encoding}
\end{figure*}





\subsection{Similar App Icon Retrieval}

During the icon retrieval process we take an app icon, calculate the required embeddings and search in the encoded space for $top-k$ nearest neighbours using either the \emph{cosine distance} or $L_2$ \emph{norm} as the distance metric.

Let $C_{target}, S'_{target}$ be respectively the content and style embedding for the icon of the \emph{target app} (i.e. the original app), for each image $i$ in the icon corpus, we calculate different distance metrics as shown in Table~\ref{Tab:DistanceMetricSuammry}. When we combine content and style embeddings, we used an empirically decided coefficient $\alpha$ that defines the relative weight between the style and content embeddings. 

\begin{table}[h]

\centering
\caption{Distance metrics} \vspace{-3mm}
\begin{tabular}{l|l}

\specialrule{.12em}{1em}{0em}
 \textbf{Distance metric} &{\textbf{Definition}} \\ \hline
 \specialrule{.12em}{0em}{0em}
\rowcolor{gray1}
\multicolumn{2}{l}{\textbf{$L_2$ norm based}} \\ \hline 
\specialrule{.12em}{0em}{0em}
$Content_{L_2}$ &  $ \norm{C_{target} - C_i}_2$ \\ \hline
$Style_{L_2}$ &   $ \norm{S'_{target} - S'_i}_2$\\ \hline
${(Content+Style)}_{L_2}$ &  $Content_{L_2}$+$\alpha$$Style_{L_2}$\\ \hline
\specialrule{.12em}{0em}{0em}
\rowcolor{gray1}
\multicolumn{2}{l}{\textbf{\emph{Cosine distance} based}} \\ \hline 
\specialrule{.12em}{0em}{0em}
$Content_{cos}$ &  $1 - \frac{C_{target}C_i}{ \norm{C_{target}}_2 \norm{C_i}_2}$ \\ \hline
$Style_{cos}$ &   $1 - \frac{S_{target}S_i}{ \norm{S_{target}}_2 \norm{S_i}_2}$ \\ \hline
${(Content+Style)}_{cos}$ & $Content_{cos}$+$\alpha$$Style_{cos}$ \\ \hline
\specialrule{.12em}{0em}{0em}
\end{tabular}
\label{Tab:DistanceMetricSuammry}
\end{table}

%
%
%

For the SIFT baseline, we only used the $L2$ distance between the SIFT vectors as defined by the original SIFT paper~\cite{SIFT}. That is, assume at the retrieval stage, for the query image $i$ we have $t_i$ descriptors and for each image $j$ in the dataset we have $t_j$ descriptors. Then, for each image  $j$  in the dataset, we find the closest pairs for all $t_i$ descriptors among $t_j$ descriptors. We define the total distance as the sum of the distances between all $t_i$ descriptors and its closest pairs.

\section{Results}
\label{Sec:Results}

\subsection{Evaluation of Embeddings}

To quantify the performance of the different embeddings we first encoded all 1.2 million app icons to the three lower dimensional embeddings we test; \emph{SIFT, content,} and \emph{style}}.  Then for each app in the {\bf \emph{labelled set}}, we retrieved the $top-5$, $top-10$, $top-15$, and $top-20$ nearest neighbours from the encoded space  for each embedding type using $L_2$ and $cosine$ distances. When we measure the distance using combined content and style embeddings, we have a hyper-parameter $\alpha$ that defines the contribution of style and the content embeddings. In cosine distance measure all the distances are between 0 and 1. However in $\mathbf{L}_2$ distance is not bounded. As such, the range of $\alpha$ differs for cosine and $\mathbf{L}_2$ distances.

The idea behind this process is that if a given embedding is a good metric for image similarity, for a given image  it must be able to retrieve all similar images in the labelled set in the nearest neighbour search. The order of the retrieval is not important as we do not know whether there are more similar images in the unlabelled corpus of app icons (e.g. the developer has multiple identities) and as we also did not rank the apps inside groups in the labelled set. We define {\bf\emph{retrieval rate}} as the percentage of labelled images we were able to retrieve out of 3,910 labelled images. One limitation of this method is that there can be scenarios where the corpus contains much better legitimate similar images than the labelled images and the number of such images is higher than the $k$ value we select. In such cases while the embeddings retrieve relevant images, those do not count towards the \emph{retrieval rate}. Thus, the results of this analysis only present the relative performance of the embeddings and a 100\% \emph{retrieval rate} is not expected. We summerise our results in Table~\ref{Tab:Results1}.


 \begin{table}
\centering
\caption{Retrieval rates for the labelled dataset} 
\resizebox{1.02\columnwidth}{!}{%
\begin{tabular}{p{4cm}|c|c|c|c}\specialrule{.12em}{1em}{0em}

{\bf Embedding } & {\bf top-5 } & {\bf top-10}  &  {\bf top-15}&  {\bf top-20}\\ \hline
\specialrule{.12em}{0em}{0em}
SIFT 								& 42.89\%	& 46.37\% & 48.51\%	& 49.45\%	\\ \hline
 \specialrule{.12em}{0em}{0em}
\rowcolor{gray1}
\multicolumn{5}{l}{\textbf{$Cosine\;distance$ based}} \\ \hline 
\specialrule{.12em}{0em}{0em}
Content$_{cos}$  					& 44.06\%	& 47.50\% & 50.51\%	& 52.40\%	\\ \hline
Style$_{cos}$ 						& 53.24\%	& 57.61\% & 61.22\%	& 63.03\%	\\ \hline
Content$_{cos}$ + $\alpha$Style$_{cos}$ \\ \hline
\ \ \ \ \ \ \ \ \ \ $\alpha=100$ 	& 53.46\%   & 57.69\% & 61.36\% & 63.13\% \\ \hline
\ \ \ \ \ \ \ \ \ \ $\alpha=10$ 	& 53.70\%   & 57.86\% & 61.71\% & 63.75\% \\ \hline
\ \ \ \ \ \ \ \ \ \ $\alpha=6$ 		& \bf{53.73\%}   & \bf{58.11\%} & \bf{61.86\%} & \bf{63.80\%} \\ \hline
\ \ \ \ \ \ \ \ \ \ $\alpha=2$ 		& 53.15\%   & 57.36\% & 61.14\% & 63.25\% \\ \hline
\ \ \ \ \ \ \ \ \ \ $\alpha=1$ 		& 51.32\%	& 55.64\% & 59.33\%	& 61.46\%	\\ \hline
\ \ \ \ \ \ \ \ \ \ $\alpha=0.5$ 	& 49.90\%	& 53.78\% & 57.11\% & 59.10\% 	\\ \hline
\ \ \ \ \ \ \ \ \ \ $\alpha=0.1$ 	& 45.83\%	& 49.67\% & 52.71\% & 54.48\%	\\ \hline

\specialrule{.12em}{0em}{0em}
\rowcolor{gray1}
\multicolumn{5}{l}{\textbf{$L_2$ norm based}} \\ \hline 
\specialrule{.12em}{0em}{0em}
Content$_{L_2}$  					&  41.32\%	&  44.59\%	& 47.36\%	&  49.08\%	\\ \hline
Style$_{L_2}$ 						&  47.76\%	&  51.39\%	&  54.46\%	&  56.24\%	\\ \hline
Content$_{L_2}$ + $\alpha$Style$_{L_2}$ \\ \hline
\ \ \ \ \ \ \ \ \ \ $\alpha= 1e6$ 	& 43.83\%	& 47.07\%   & 49.95\%   & 51.57\% \\ \hline
\ \ \ \ \ \ \ \ \ \ $\alpha= 1e7$ 	& 47.72\%	& 51.57\%   & 54.87\%   & 56.74\% \\ \hline
\ \ \ \ \ \ \ \ \ \ $\alpha= 1e8$ 	& 48.66\%	& 52.09\%	& 55.40\%	& 56.93\% \\ \hline
\ \ \ \ \ \ \ \ \ \ $\alpha= 1e9$ 	& 47.75\%	& 51.48\%	& 54.57\%	& 56.28\% \\ \hline

\specialrule{.12em}{0em}{0em}
\end{tabular}}

\label{Tab:Results1}
\end{table}

 \begin{figure*}
\centering
\includegraphics[width=.7\textwidth]{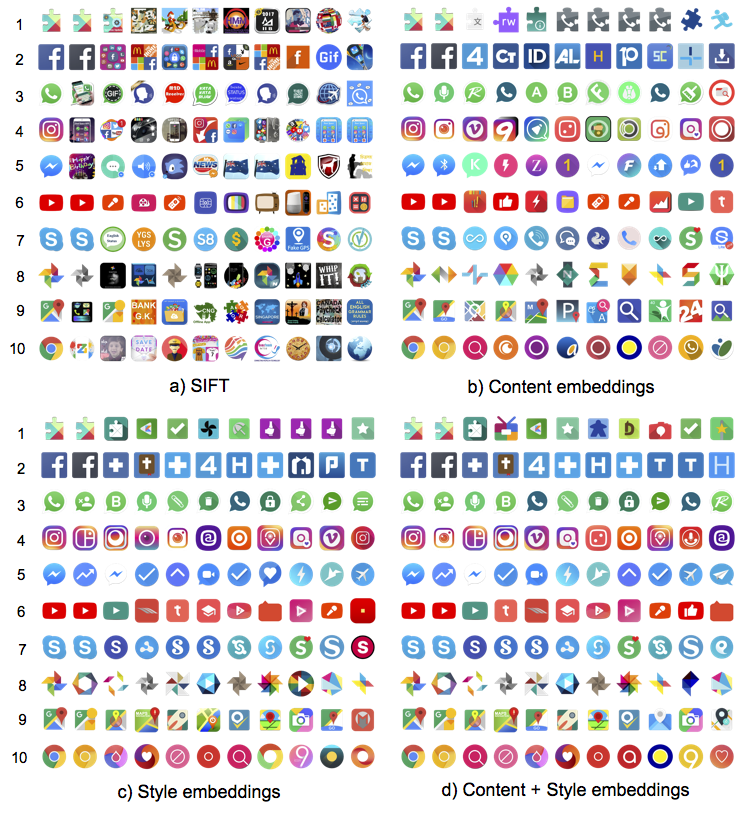}%
\caption{Top-10 visually similar apps for the 10 top ranked apps}
\label{Fig:top_10_embeddings}
\end{figure*}

According to the results in Table~\ref{Tab:Results1}, neural embeddings perform better than the SIFT features. Also, it is noticeable that \emph{style embeddings} outperforms content embeddings. For example, for four top-k scenarios we consider the style embeddings have approximately 8\% -- 9\% higher retrieval rates than the content embeddings for cosine similarity and  an approximately 6\% higher retrieval rate for $L_2$ distance. Results also show that cosine distance produces better results in majority of the scenarios. Combining stye embeddings with content embeddings increases the performance slightly. For example, at best case ($\alpha = 6$; i.e. style embeddings carry six times weight compared to content embeddings) combining embeddings increase the retrieval rate by approximately 0.5\%--0.6\%.

To elaborate further on the performance of the embeddings qualitatively, In \figurename~\ref{Fig:top_10_embeddings} we show the top-10 visually similar apps we retrieved under different embeddings for some well-known apps. \figurename~\ref{Fig:top_10_embeddings}-(a) shows that apart from identifying 1-2 similar apps (E.g. row 1 - Google Play Services, row 2 - Facebook, and row 7 - Skype), SIFT does not identify visually similar apps. Results of content embeddings shown in \figurename~\ref{Fig:top_10_embeddings}-(b) indicate that they perform better and has identified good fits in several cases (E.g. row 1 - Google Play Services and E.g. row 9 - Google Maps). The improvement provided by style embeddings is visible in some of the cases in \figurename~\ref{Fig:top_10_embeddings}-(c). For instance, style embeddings has retrieved app icons that have the same \emph{``look and feel''} in terms of colour for Facebook Messenger (row 5) and Skype (row 7). Finally, the combined embeddings haven't perform significantly different from style embeddings alone as can be seen in \figurename~\ref{Fig:top_10_embeddings}-(d). This can be expected as there was only approximately 1\% improvement in combined embeddings in the labelled set.

\subsection{Retrieving Potential Counterfeits}

We next use the embeddings that performed best (Content$_{cos}$ + $\alpha$Style$_{cos}$ where $\alpha=6$) to retrieve visually similar apps for {\bf \emph{top-1,000}} apps and check the availability of malware, as spreading malware is one of the main objectives behind publishing counterfeit apps. In this analysis, we focus only on the top apps since they usually are the main targets of counterfeits. For each app in  {\bf \emph{top-1,000}} we retrieved $top-10$ visually similar apps from  the corpus of 1.2 million apps that are not from the same developer and with the same category as the top app. However, the $top-10$ neighbour search is forced to return $10$ closest app icons irrespective of the distance distribution and as a result there can be cases where the search returns 10 results, nonetheless they are a large distant apart from the original app. Thus, we applied a distance threshold to further narrow down the retrieved results. We varied the distance threshold from 0 to 1 at 0.01 intervals and calculated the retrieval rate in the labelled set for each threshold value (i.e. percentage of retrieved apps within the radius of the threshold). Then we calculated the knee-point~\cite{satopaa2011finding} of the retrieval rate against the threshold graph, after which the improvement of retrieval rate is not significant. The threshold value we calculated was 0.27 (note that is the normalised distance as for the embedding considered here, the closet possible distance is zero while the maximum possible distance apart is 7). 

This process resulted 6,880 unique apps that are potentially counterfeits of one or more apps with in top-1,000. Out of this 6,880 we had APK files for 6,286 apps. In \figurename~\ref{Fig:graph}, we show a graph-based visualisation of the app icons of potential counterfeits we identified for top-100 apps. The centre node of each small cluster represent an app in top-100 and the connected apps to that are the visually similar apps we identified for that particular app.

\begin{figure}
\centering
\includegraphics[width=.5\textwidth]{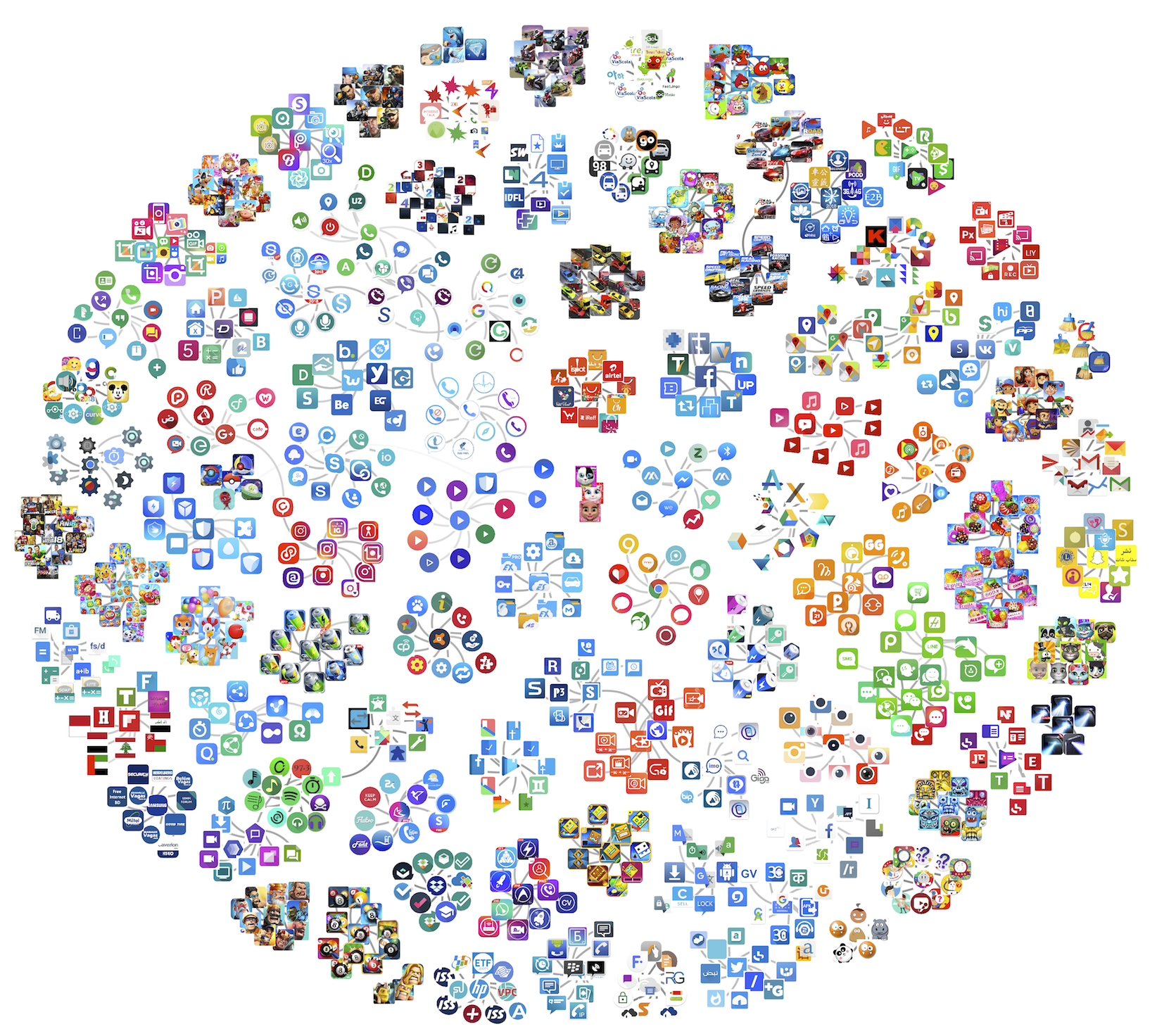}%
\caption{Graph based visualisation of top-100 apps and visually similar app icons (Small clusters in this figure do not contain apps from the same developer)}
\label{Fig:graph}
\end{figure}

\subsection{Malware analysis}

We then checked each of the 6,286 potential counterfeits using the private API of the online malware analysis tool \emph{VirusTotal}. VirusTotal scans the APKs with over 60 commercial anti-virus tools (AV-tools) in the likes of AVG, Avast, Microsoft, BitDefender, Kaspersky, and  McAfee and provides a report on how many of the tools identified whether the submitted APKs contain malware.  In \figurename~\ref{Fig:MalwareCount} we show a summarised view of the number of apps that were tagged for possible inclusion of malware by one or more AV-tools in VirusTotal and their availability in Google Play Store as of \emph{19-04-2018}. As the figure shows there are 853 APKs that are tagged by at least one of the AV-tools for possible inclusion of malware.  

However, there can be false positives and as such a single AV-tools tagging an APK as malware in VirusTotal may not necessarily mean that the APK contains malware. As a result, previous work used different thresholds for the number of AV-tools that must report to consider an APK as malware. Ikram et al.~\cite{ikram2016analysis} used a conservative threshold of 5 and Arp et al.~\cite{arp2014drebin} used a more relaxed threshold of 2. \figurename~\ref{Fig:MalwareCount} shows that we have 340 apps if the  AV-tool threshold is 2 and 139 apps if the threshold is 5, out of which 305 and 122 apps respectively, are still there in Google Play Store. 

\begin{figure}
\centering
\includegraphics[width=.5\textwidth]{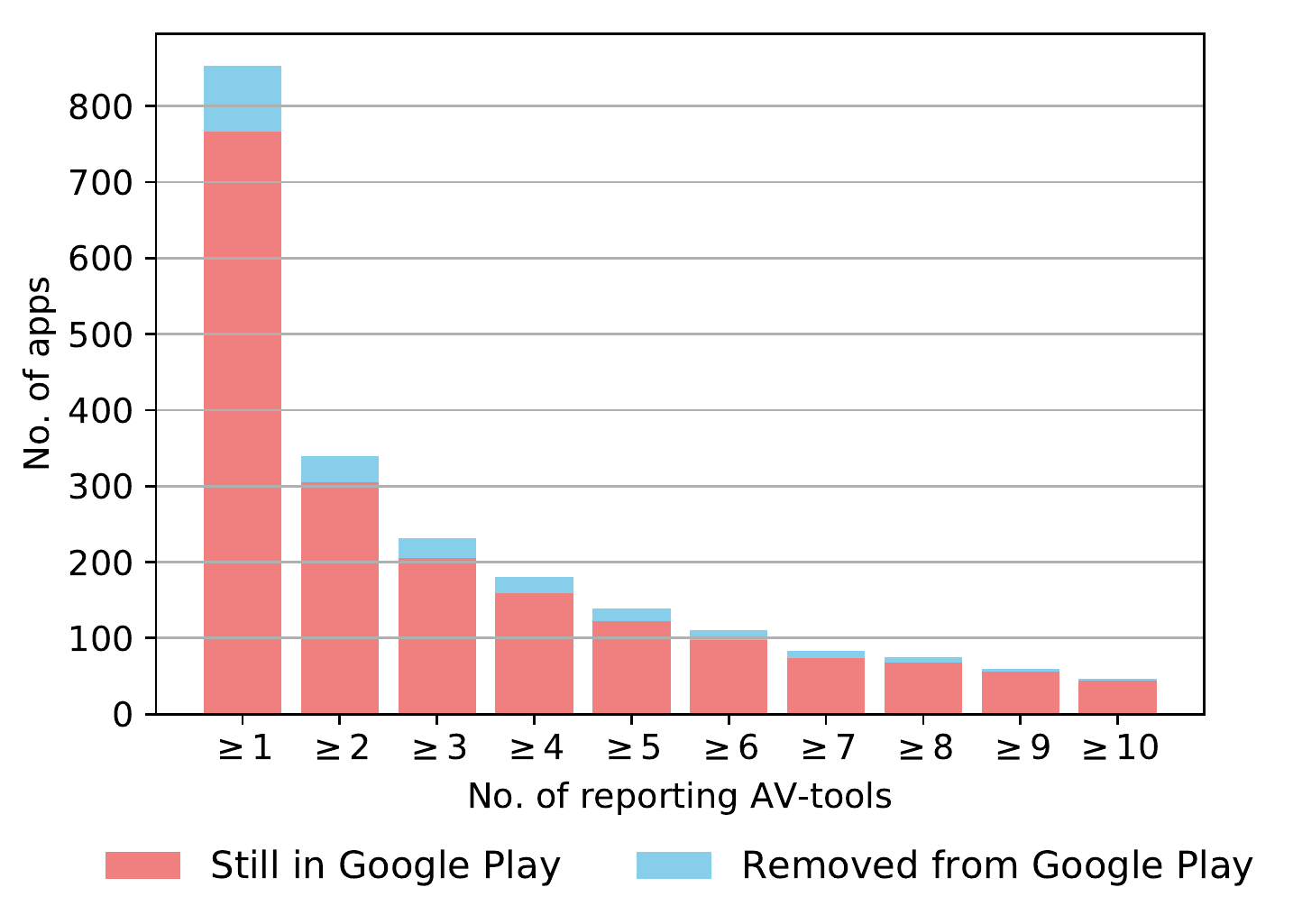}%
\caption{Number of apps against  the number of reporting AV-tools in VirusTotal}
\label{Fig:MalwareCount}
\end{figure}

In Table~\ref{Tab:ExampleMalware}, we show some example apps that were tagged as containing malware and their original app. We also show the number of downloads, number of requested permissions, and the number embedded advertisement libraries obtained by decompiling the app and using a previously published list of advertisement libraries~\cite{seneviratne2015measurement} for both the original app and the potential counterfeit. The table shows that while the counterfeit does not attract as large numbers of downloads as the original app in some occasions they have been downloaded significant number of times (e.g. Temple Theft Run). In an extreme case, a potential counterfeit for \emph{DU Battery Save} asks approximately 9 times more permissions than the original app.

\begin{table*}
\scriptsize
\centering
\caption{Example visually similar apps that contain malware}
  \begin{tabular}{cccccccccc}
    \specialrule{.12em}{0em}{0em}
    \multicolumn{2}{c}{{\bf App icon}} &
    \multirow{2}{*}{\bf AV Tools} &
      \multicolumn{2}{c}{\bf Downloads} &
      \multicolumn{2}{c}{\bf Permissions} &
      \multicolumn{2}{c}{\bf Ad. libraries} \\
    {Original} & {Similar} &  & {Original} & {Similar} & {Original} & {Similar} & {Original} & {Similar} \\  \\  
      \specialrule{.12em}{0em}{0em} \\ 
      
        \begin{minipage}{.2\textwidth}
        Temple Run 2
        \centering 
      \includegraphics[scale=0.08]{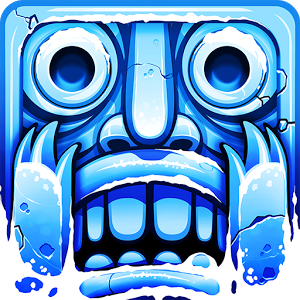}
    \end{minipage}  &           \begin{minipage}{.2\textwidth} \centering Temple Theft Run*
      \includegraphics[scale=0.08]{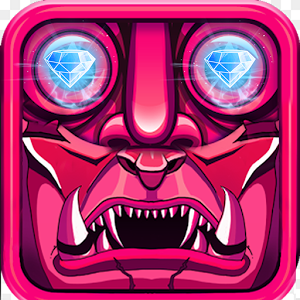}
    \end{minipage} & 6 & 500M+ & 500k+ & 8 & 4 & 6 & 10 \\ \vspace{-2mm} \\ \hline \\  
    
            \begin{minipage}{.2\textwidth}
            \centering
        Minecraft$\dagger$ \\
        \centering 
      \includegraphics[scale=0.08]{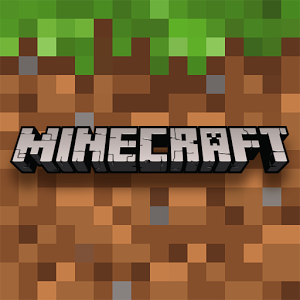}
    \end{minipage}  &           \begin{minipage}{.2\textwidth} \centering Survival Blocks
      \includegraphics[scale=0.08]{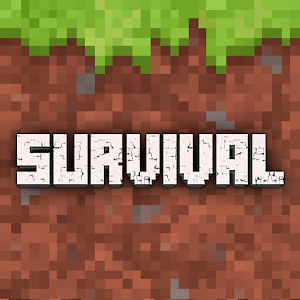}
    \end{minipage} & 5 & 10M+ & 10k+ & - & 5 & - & 6 \\ \vspace{-2mm} \\ \hline \\

            \begin{minipage}{.2\textwidth}
            \centering
        Parallel Space \\
        \centering 
      \includegraphics[scale=0.08]{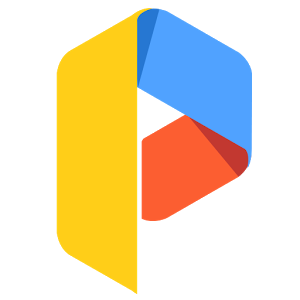}
    \end{minipage}  &           \begin{minipage}{.2\textwidth} \centering Double Account
      \includegraphics[scale=0.08]{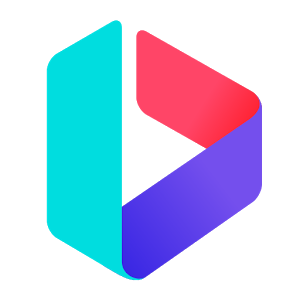}
    \end{minipage} & 15 & 50M+ & 100k+ & 130 & 145 & 3 & 2 \\ \vspace{-2mm} \\ \hline \\  
    
            \begin{minipage}{.2\textwidth}
            \centering
        Sniper 3D \\
        \centering 
      \includegraphics[scale=0.08]{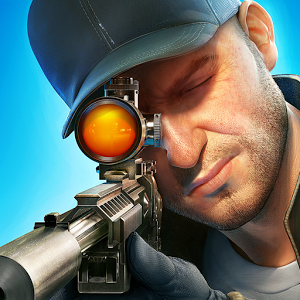}
    \end{minipage}  &           \begin{minipage}{.2\textwidth} \centering Elite Sniper Killer*
      \includegraphics[scale=0.08]{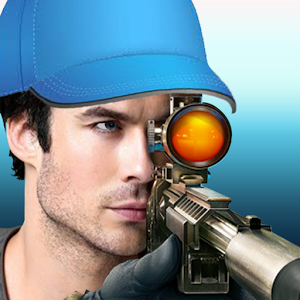}
    \end{minipage} & 6 & 100M+ & 500+ & 11 & 7 & 11 & 6 \\ \vspace{-2mm} \\ \hline \\  
    
                \begin{minipage}{.2\textwidth}
            \centering
        Clean Master \\
        \centering 
      \includegraphics[scale=0.08]{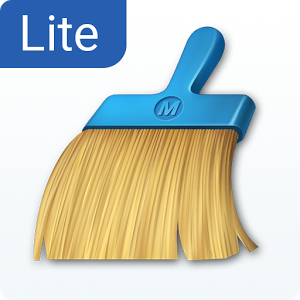}
    \end{minipage}  &           \begin{minipage}{.2\textwidth} \centering RAM Booster*
      \includegraphics[scale=0.08]{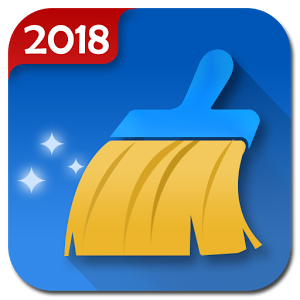}
    \end{minipage} & 5 & 50M+ & 500+ & 34 & 22 & 4 & 1 \\ \vspace{-2mm} \\ \hline \\
    
                    \begin{minipage}{.2\textwidth}
            \centering
        DU Battery Save \\
        \centering 
      \includegraphics[scale=0.08]{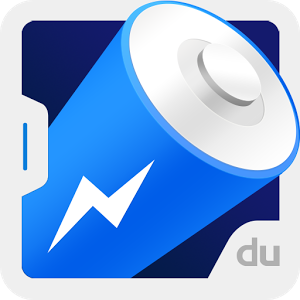}
    \end{minipage}  &           \begin{minipage}{.2\textwidth} \centering Battery Saver
      \includegraphics[scale=0.08]{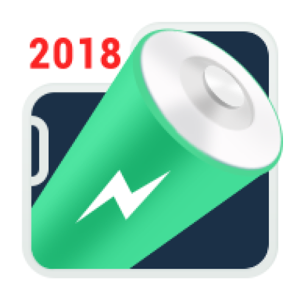}
    \end{minipage} & 6 & 100M+ & 1k+ & 24 & 213 & 3 & 2 \\ \vspace{-2mm} \\ 
    
    \specialrule{.12em}{0em}{0em}
  \end{tabular}
   \newline * App is currently not available in Google Play Store \\  $\dagger$ APK is not available as the app is a paid app
   \label{Tab:ExampleMalware}
\end{table*}

\section{Discussion}
\label{Sec:Discussion}

Using a large dataset of over 1.2 million app icons and over 1 million app executables, in this paper we presented insights to the app counterfeit problem in mobile app markets. The objective of the proposed embeddings-based method is to quickly and automatically assess a new submission and decide whether it resembles an existing app. If the new app is visually similar to an existing app, the app market operator can decide to do further security checks that can potentially include dynamic analysis as well as manual checks. We next discuss the limitations of our work and possible future extensions.

\subsection{Limitations}

Establishing a ground truth dataset for this type of problem is challenging due to several reasons. In this work, to build the ground truth dataset we used a heuristic approach to shortlist groups of apps that can potentially show visual similarities and then refine the dataset by manual inspection. However, as described in Section~\ref{Sec:Results}, there is a possibility that the unlabelled portion of data can still contain better similar apps than the labelled set and such apps can be returned during the nearest neighbour search instead of the target apps. This will result in a lower performance in terms of \emph{retrieval rate}; yet in reality received images also show high visual similarity. One possible solution for this is to use crowdsourcing to directly evaluate the performance of the embeddings without using a labelled dataset. For instance, retrieved images can be shown to a set of reviewers together with the original image and ask them to assign values for similarity with the original image. Then these values can be aggregated to come up with an overall score for the performance of the embedding. Crowdsourcing will also alleviate any biases introduced by individuals as visually similarity of images in some occasions can be subjective.

\subsection{VGGNet Fine-tuning}

In this work we used a pre-trained VGGNet to generate both content and style embeddings. However, the VGGNet was trained for content identification and further fine tuning of VGGNet could improve the style embeddings, which in return may increase retrieval rates. For example, one such approach may be to use a different loss function such as perceptual loss~\cite{johnson2016perceptual} compared to the current cross entropy loss function and fine-tune the VGGNet so that it can further capture the similarities in styles in addition to the content. Another possible approach is to use \emph{triplet loss} to train the network~\cite{wang2014learning}. That is, at training time the network is shown sets of three images; base image, a highly visually similar image to the base image, and  a highly dissimilar image to the base image.

Another potential improvement is the fusion of activations of several convolutional layers. In this work we used only the fifth convolution layer of the VGGNet to generate the style embeddings. This can be further explored to check whether creating embeddings from several convolution layers and combining them in some form (e.g. linear weighted sums or deep features using autoencoders) will improve the retrieval rates.

\begin{figure}
\centering
\includegraphics[scale=0.15]{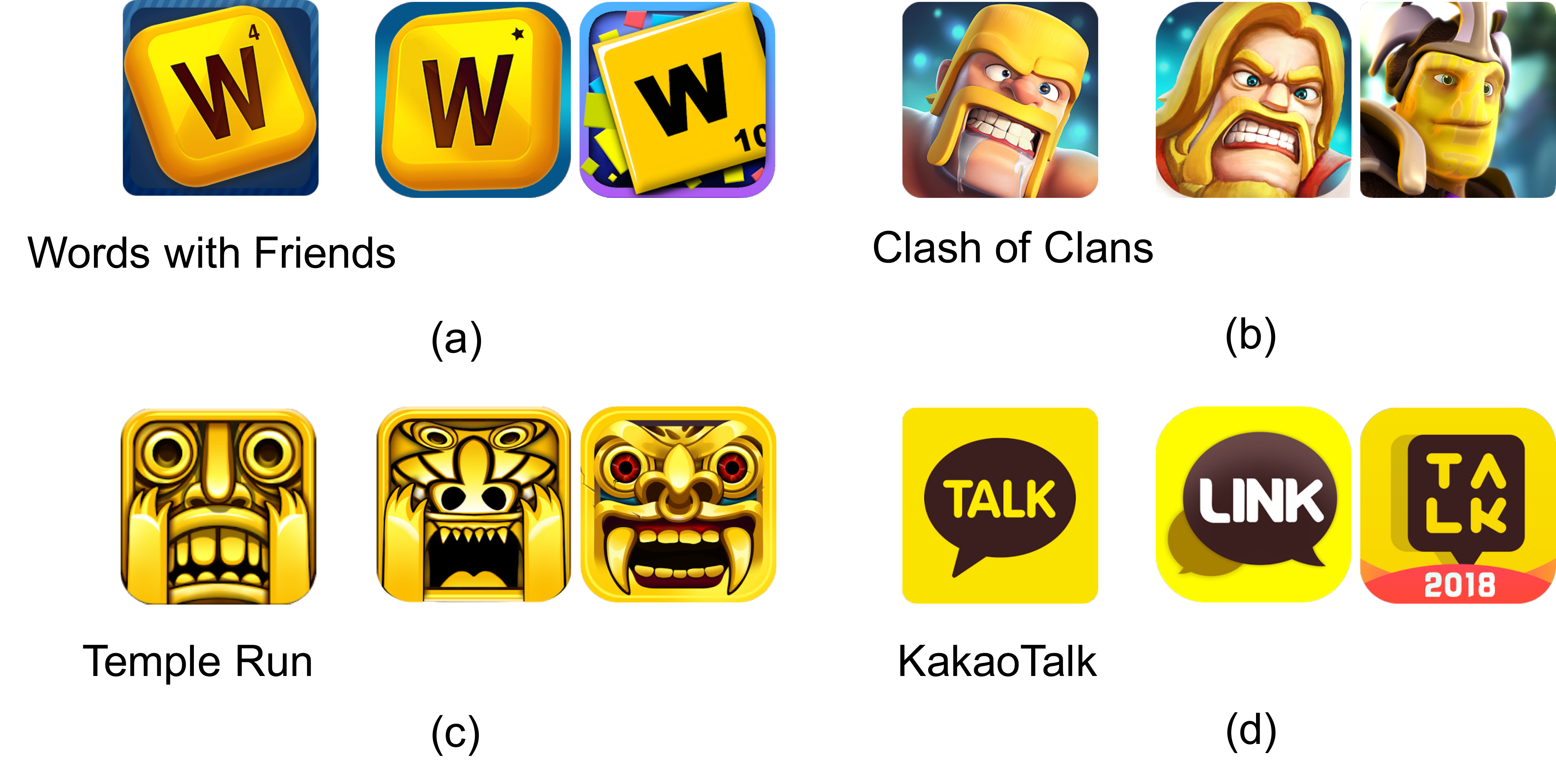}%
\caption{Some example apps that showed high visual similarity to one of the apps in top-1,000, yet did not contain any malware}
\label{Fig:NonMalware}
\end{figure}

\subsection{Identifying Counterfeits}

Our evaluation of the retrieved set of highly visually similar apps was limited to the apps that possibly contain malware. Nonetheless, there can be counterfeits that do not contain malware and sometimes it can be difficult to  automatically decide whether a given app is counterfeit or not. In \figurename~\ref{Fig:NonMalware} we show some examples we retrieved that showed high visually similarity to one of the apps in top-1,000 yet did not contain malware or showed a significant difference in permissions or ad libraries. For instance, in \figurename~\ref{Fig:NonMalware}-(a) we show two visually similar apps we retrieved for  the popular game \emph{Words with Friends}. While the icons are similar it is difficult to decide whether these are true counterfeits and as a result some other factors such as app functionality and description need to be considered. 

For such scenarios, instead of the using only the similarity in app icons, the overall similarity of all the images available in Google Play Store pertaining to the two apps can be considered. This is because a developer can make the icons slightly different from the original app, and yet have the same visual \emph{``look and feel''} inside the app. Also, a number of work highlighted that apps can be clustered based on functional similarity using text mining methods~\cite{gorla2014checking,surian2017app}. Combining such methods with state-of-art techniques such as \emph{word vectors} and \emph{document vectors}~\cite{mikolov2013efficient,le2014distributed} and using them in conjunction with image similarity can further improve results. Nonetheless, for some cases still a manual intervention may be required. For example, in above case of \emph{Words with Friends} the similar apps are also word games and they are likely to show high visual similarity in internal GUIs as well as textual descriptions. In such scenarios again it might be useful to rely on crowdsourcing to obtain an overall assessment.
\section{Conclusion}
\label{Sec:Conclusion}

Using a dataset of over 1.2 million app icons and their executables, we presented the problem of counterfeits in mobile app markets. We proposed an icon encoding method that allows to efficiently search potential counterfeits to a given app, using neural embeddings generated by a state-of-the-art convolutional neural network. More specifically for app counterfeit detection problem, we showed that \emph{style embeddings} generated by the Gram matrix of the $conv5\_1$ convolution layer of a pre-trained VGGNet is more suitbale compared to commonly used content embeddings (generated by the last fully connected layer - $fc\_7$) and SIFT baselines. Our results also indicated that combining content and style embeddings can further increase the detection rates of counterfeits. For the top-1,000 apps in Google Play Store, we identified a set of 6,880 highly visually similar apps and we found that under a highly conservative assumption, 139 of them contain malware. To the best of our knowledge this is the first work to do a large-scale study  on the counterfeits in mobile app markets using image similarity methods and propose effective embeddings to automatically identify potential counterfeits. \\

\section*{Acknowledgments}
 
Authors would like to thank VirusTotal for kindly providing access to the private API, which was used for the malware analysis in this paper.

\bibliographystyle{IEEEtran}
\bibliography{IEEEabrv,bibliography.bib}

\end{document}